# A Coverage Strategy for Wireless Sensor Networks in a Three-dimensional Environment

## Lin Feng[1], Tie Qiu[2], Zhenlong Sun[1], Feng Xia[2]*, Yu Zhou[1]

1 School of Innovation Experiment, Dalian University of Technology, Dalian 116024, China
2 School of Software, Dalian University of Technology, Dalian 116620, China
E-mail: fenglin@dlut.edu.cn
E-mail: qiutie@dlut.edu.cn
E-mail: txle@mail.dlut.edu.cn
E-mail: f.xia@ieee.org
E-mail: zhouyujoe@qq.com
*Corresponding author

**Abstract**: Coverage is one of the fundamental issues in wireless sensor networks (WSNs). It reflects the ability of WSNs to detect the fields of interest. In a real sensor networks application, the detection area is always non-ideal and the terrain of the detection area is often more complex in applications of three-dimensional sensor networks. Consequently, many of the existing coverage strategies cannot be directly applied to three-dimensional spaces. This paper presents a new coverage strategy for the three-dimensional sensor networks. Sensor nodes are uniformly distributed. The cost factor is utilized to construct the perceived probability and the classical watershed algorithm after the transformation of points from the three-dimensional space to the two-dimensional plane using the dimensionality reduction method, which can maintain the topology characteristic of the non-linear terrain. The detection probability in the optimal breath path is used as the measure to evaluate the coverage. Simulation results indicate that the proposed strategy can determine the coverage with fewer nodes, while achieving the coverage requirements of the networks.

**Reference** to this paper should be made as follows:

**Biographical notes**: Lin Feng is a Professor and Ph.D Supervisor in School of Innovation Experiment, Dalian University of Technology, China. His research interests cover date mining, wireless sensor networks and Internet of Things.

Tie Qiu received his PhD in computer science and technology from Dalian University of Technology, China. He is a Lecturer in School of Software, Dalian University of Technology. His research interests cover embedded systems, Internet of things and systems modelling. He is a member of China Computer Federation and ACM.

Zhenlong Sun received B.E. degree from Dalian University of Technology, China. Currently he is a Master student in School of Innovation Experiment, Dalian University of Technology. His research interest covers wireless sensor networks.

Feng Xia is an Associate Professor and PhD Supervisor in School of Software, Dalian University of Technology, China. He is the (Guest) Editor of several international journals. He serves as General Chair, PC Chair, Workshop Chair, Publicity Chair, or PC Member of a number of conferences. Dr. Xia has authored/co-authored one book and over 130 papers. His research interests include social computing, network science, mobile computing, and cyber-physical systems. He is a Senior Member of IEEE and a member of ACM.

Yu Zhou received B.E. degree from Dalian University of Technology, China. Currently he is a Master student in School of Innovation Experiment, Dalian University of Technology. His research interest covers Internet of Things.

# 1 Introduction

A wireless sensor network (WSN) is a "smart" system, which consists of numerous small, low-powered, self-organizing sensor nodes that have communication and computation capabilities. The sensor nodes of networks can be used to complete tasks assigned according to the application environment. WSNs have been widely used (Akyildiz et al., 2002) in applications related to the military, environmental monitoring, health care, and so on. However, the computation power and communication capacity of sensor nodes within the network are often limited by the environment. The sensor network always has some special characteristics such as the dense distribution of sensor nodes, frequently changing topology, multihop communication mode, and so on. To manage the sensor network more effectively and ensure a better quality of network service, we should consider the coverage issue of networks, which refers to the method of deployment of nodes to achieve better detection. Coverage reflects networks' "perceived quality of service" and provides a more reliable guarantee to monitor and control the sensor networks.

The coverage problem, which involves the appropriate placing of the sensor nodes in the surveillance area to meet the networks' coverage requirements, is one of the fundamental issues in WSNs. Many researchers have studied this issue and achieved great results, such as the coverage in two-dimensional (2D) plane or the full three-dimensional (3D) space. In studies of the coverage problem in 3D space (Huang et al., 2004; Watfa and Commuri, 2006), the 3D space is often considered to be ideal. However, in real-world applications, the surveillance area is often a complex surface. Many studies (Zhao et al., 2007; O'Rourke, 1992) have proved that the deterministic sensor-deployment problem is nondeterministic polynomial-time complete (NP-complete), that is, it cannot be solved in polynomial time and only has an approximate solution. The applications of WSNs are not in a 2D plane or a 3D space but on a complex surface topography, and they are therefore affected by the surface topography. The previously proposed coverage models related to the 3D space cannot completely reflect the coverage ratio so that the previous 2D place or 3D space deployment strategies cannot be used to deal with complex surfaces.

Considering these facts, the manner of deployment of sensor nodes to cover the entire monitored area is essential to build energy-efficient and self-organizing sensor networks. In this study, we propose a new solution for the complex 3D surfaces using the manifold learning algorithm for the well-known nonlinear dimensionality reduction, called locally linear embedding (*LLE*), to convert the 3D space into a 2D space. Then we determine the figures of height related to the detection probability in the 2D plane according to the parameters defined below after applying the perceived probability model to the coverage strategy. Next, we obtain the water contours using the watershed algorithm of the image-processing method and propose an improved algorithm based on the shortest path to find an optimal breath path along which the mobile object can pass through the surveillance area. The perceived probability in this path is regarded as a parameter that represents the coverage. If the optimal perceived probability in this path can meet the requirements of the network, the others can too.

The remainder of this paper is divided into the following sections. Section 2 provides a summary of previous research on coverage strategies. Section 3 proposes a new coverage strategy appropriate for the 3D surface and elaborates this method in details for real world applications related to WSNs. Section 4 proposes an improved algorithm based on the shortest path to find an optimal breath path. Section 5 describes the experiments, simulates the strategy, and analyzes the results. Section 6 concludes the paper.

# 2 Background

## 2.1 Related Work

Numerous researchers have carried out research on the coverage issue for 2D detection areas. Meanwhile, the coverage strategies in 2D space have been widely applied in the case of WSNs. For example, some studies (Hoffmann et al., 1991; Shermer, 1992) describe the gallery problem in a 2D space, which aims to arrange the least number of guarders to ensure that every point in this polygonal area is observed by at least one guard. Another study (Dhillon and Chakrabarty, 2003) places the sensor nodes based on a deterministic approach and proposes the deployment algorithm depending on a greedy method. The authors of this study choose grid points to place the sensor nodes until all the nodes could satisfy the constraints each time. However, this methodology is not suitable for applications in complex and dangerous environments prevailing in real world. Other studies (Liu et al., 2005; Lazos and Poovendran, 2006) refer to a strategy that places the sensor nodes randomly. In the study by Howard et al. (2002), the sensor nodes are regarded as virtual charged particles with the ability to move, and the sensor nodes affected by a force do not stop moving until all nodes are in equilibrium. Some researchers (Alam and Haas, 2006) have studied not only the coverage problem but also the connectivity issue. The optimal geographical density control (OGDC) algorithm (Zhang and Hou, 2004) maintains the coverage and connectivity in large-scale WSNs. Some studies (Wang et al., 2005) deal with the coverage and connectivity for an arbitrarily shaped model without using any previous model. Huang and Tseng (2005) propose the concept of circular coverage, assuming that the detection area can be fully covered if the detection circles of all the sensor nodes are fully covered. Huang and Tseng (Huang et al., 2004) consider the coverage issue in the context of transformation from the 2D circle to the 3D sphere and settle a 3D-coverage problem using a distributed approach without increasing the computational complexity.

Some researchers have recently begun the study of coverage strategies in 3D space. For instance, in some actual scenarios, the sensor nodes can be placed anywhere within the 3D volume. The study by Alam and Haas (2006) analyzes grid points' coverage and connectivity issues, such as the need to deploy sensor nodes in underwater marine areas to detect a range of oceanic environments. Bai et al. (2009) study a method to achieve the network connectivity with the premise of a complete coverage. The authors in another report (Oktug et al., 2008) have studied the coverage problem on the surface of a 3D topography. In some studies (Zhao et al., 2009; Liu and Ma, 2011), the authors deeply explore the expected coverage ratio, which also belongs to the surface coverage, by considering different 3D terrain models from different aspects. Some researchers (Jin and Rong, 2012) have also studied the surface coverage problem which is related to the optimal sensor deployment on the three-dimensional terrains to achieve the highest overall sensing quality according to the actual application environment. For the optimal coverage control problem, a general function is introduced to measure the unreliability of the monitored data in the entire sensor networks including the perception ability of sensor nodes determined by the distances of nodes. However, the coverage holes of the entire sensor networks have not been resolved well.

*2.2 Problem Statement*

The 3D detection area is ideal according to the above discussion, so the previous deployment strategies can be only applied to the full space. In general, considering that the monitoring regions are often rough surfaces, sensor nodes are deployed not at arbitrary points but on the surfaces, such as volcanoes, canyons and other hilly geographical areas, of the region. Previous coverage strategies may thus lead to the "empty holes" problem (Zhao et al., 2009). As shown in Figure 1, some areas fall within the detection-blind region of nodes, which means that the 2D-coverage strategies cannot be directly applied to an actual 3D detection space. Under these circumstances, the strategy of transformation from 2D to 3D space is often not very easy to achieve complete coverage for such problems. Given that the sensor nodes are directly placed in the 3D space, the locations of sensor nodes are determined in advance if the nodes are planned to be distributed manually or by a robot. This problem has been proved to be NP-complete, which cannot be solved in polynomial time. Moreover, the nodes have uncertain positions after random deployment. Then the sensor coverage strategy proposed in the study by Watfa and Commuri (2006) cannot be applied to this 3D space either. Furthermore, this strategy not only needs more sensor nodes but also accelerates the consumption of energy within the networks. A direct analysis of the coverage strategy using the 3D model is often more complex and not very easy to understand after establishing the actual detection environment. Thus accomplishing optimal coverage more effectively with lesser number of sensor nodes is essential to reduce the energy consumption, prolong network lifetime, and improve the stability and reliability of the entire sensor networks. In this study, the deployment depends on the actual applications and we do not undertake an analysis of direct deployment on the 3D terrain model. A new strategy that can maintain the characteristics of the 3D terrain and the topology of the data points is introduced to reduce the space dimensionality. Finally, we analyze the deployment coverage of the sensors.

**Figure 1** Sensor nodes distributed on a regular surface

## 3 Detection models and deployment strategy

We propose a new solution for the transformation of a 3D surface to a 2D plane using the manifold learning algorithm for the nonlinear dimensionality reduction. Then we determine the figures of height and obtain the water contour according to the cost value and the cost factor. Finally, we propose an improved algorithm based on the shortest path to find the optimal breath path.

*3.1 3D-simulation terrains*

Previous studies (Bai et al., 2009) on coverage have mainly focused on ideal models, such as full space, wherein the sensor nodes can be placed anywhere (Alam and Haas, 2006). However, the actual applications of WSNs are often in complex terrains, which may be variable, such as volcanoes, canyons, basins, and so on, as shown in Figure 2.

**Figure 2** Actual applications: (a) is the real world terrain, and (b) is the simplified regular surface

To construct the actual applications more visually, as shown in Figure 2, the surface in (a) can be simplified to the regular and continuous surface map similar to that in (b). We use multiple peaks that are subject to Gaussian distribution to simulate the actual applications and further assume that the topographic terrain built on the 3D space is regular, nonlinear, and continuous.

In the following analysis, we assume that there are $L \times W$ grid points on the 3D multi-peak surface topography with the reflection of the real-world terrain in a range of $L \times W$ $m^2$ and $L$ is the length and $W$ is the width and the surface of peak is defined by

$$h(n) = \omega \times normcdf(peak(n), 0, 1) \qquad (1)$$

$n=1, 2 \ldots K$, where $K$ is the number of peak points that develop a peak graph respectively within the detection area. The coordinates $(i, j, h(n))$ represent one of the $K$ peak maps, where $i$ ranges from 1 to $L$ and $j$ ranges from 1 to $W$. The function *normcdf* represents the cumulative distribution function (Devore, 2011) subject to the Gaussian distribution with zero mean and unit variance, which means that the probability is less than the value in the statistics and probability. In this study, as shown in Figure 3, we construct the 3D peak maps using the function. $(x(n), y(n))$ are the coordinates of the peak points and both ρ and η are the peak parameters. The peak coefficient ω determines the height of each peak, where $e(n) = 1/sqrt\left(\left(i-x(n)\right)^2 + \left(j-y(n)\right)^2\right)^\eta$, and $peak(n) = sqrt(\rho \times e(n))$.

**Figure 3** Peak map ($K = 1$, $\omega = 20$)

In this study, we have constructed the 3D terrain with the available $K$ ranging from 1 to 50, using appropriate peak parameters, peak coefficients, and all peak points so that the terminal terrain is composed of $K$ peak maps represented as follows:

$$Height(i,j) = \max_{\{n=1,2..K\}} h(n) \qquad (2)$$

$(i, j, Height(i,j))$ are the coordinates of the grid points on the 3D terrain shown in Figure 4.

**Figure 4** Three-dimensional terrain ($K=20$)

There are 2500 data points in Figure 4, but for an actual terrain, the locations of sensor nodes have been determined when the nodes have been planned to be distributed, which has been proved NP-complete. Even though the nodes are placed randomly, the sensor-coverage strategy in the study by Watfa and Commuri (2006) cannot be applied to the 3D terrain.

If the sensor nodes are placed in point *A* in Figure 1, a part of regions limited by the 3D terrain will lie in the detection-blind area of nodes and can thus be ignored, which may lead to the "empty holes" problem.

*3.2 Transformation from 3D to 2D space*

As described previously, the coverage strategy in a 2D plane or the 3D full space cannot be directly applied to actual applications. Thus we introduce the LLE (Locally linear Embedding) algorithm to complete the transformation from a 3D space to a 2D plane. The LLE concept (Roweis and Saul, 2000) is a recently proposed nonlinear dimensionality reduction method that enables the 2D data points to maintain the original topology of the 3D points, which is suitable for the non-linear manifold. It considers the manifold to be linear locally and uses the local symmetry of the linear reconstruction to identify the neighbour relationships of high-dimensional data points so as to map the high-dimensional points to low-dimensional space.

According to the sample points, the LLE algorithm can calculate each sample point's $k$ neighbors which are $k$ sample points with the nearest distance. The reduction parameter $k$ is assigned a value that defines the process of finding the $k$ neighbors of each sample point. An appropriate parameter $k$ can make the transformation process from 3D to 2D space better. In addition, the local reconstruction weight matrix of one sample point can be calculated from all its neighbour points. Then this sample point and its weight matrix will lead to the corresponding sample points which are mapped from high-dimensional space to low-dimensional space.

**Figure 5** Comparison between PCA and LLE

In Figure 5, the results of dimensionality reduction by PCA (Principal Component Analysis) which is fit for the linear manifold and LLE are compared. LLE can achieve a better two-dimensional plane.

The 3D terrain constructed in this study is regular and nonlinear. The LLE algorithm yields $L \times W$ grid points in the 2D plane by reduction from the 3D space in the detection area. With reference to the 3D surveillance area, we should ensure a topology that is not only better, but also truly reflects the terrain characteristics of the 3D surface, in addition to avoiding possible blind detection problems. Therefore, we require a reduction parameter, which is named the cost value $\lambda$ that is defined by

$$\lambda = \begin{cases} 0 & (i = j) \\ d_{ij} / |h_i - h_j| & (i \neq j) \end{cases} \qquad (3)$$

As shown in Figure 6, $|h_i - h_j|$ is the difference in the altitudes between data points before using the LLE, whereas $d_{ij}$ in Figure 7 is the Euclidean distance between the points $i$ and $j$ in the 2D plane and $h_i \neq h_j$. After distributing the sensor nodes, the computational time will become longer if the distances between the points in our perceived probability model are calculated using the dijkstra shortest-distance algorithm. The neighboring weights of each point in LLE remain unchanged during the translation, rotation, stretching transformation, which is called no deformation of the popular learning. This characteristic ensures the conversion from the 3D space to 2D plane by the parameter $\lambda$, adding the topological characteristics of the area to be covered in WSNs.

When the topological invariant is taken into consideration, $d_{ij}$ can replace the 3D distance of the data points and the cost factor $\lambda$ can reflect the 3D surface features and the topology relationship between the data points.

**Figure 6** Altitude difference of data points on 3D terrains

**Figure 7** Euclidean distance between the 2D points

We can see that, by using the LLE algorithm, four points in Figure 6 are mapped into Figure 7. $|h_1 - h_2|$ is the altitude difference between the data points 1 and 2. Here $d_{12}$ is the

Euclidean distance between points 1 and 2 in Figure 7. The parameter $\lambda$, which reflects the 3D terrain, is determined by both the altitude difference and the distance.

*3.3 Perceived Probability*

We can obtain the corresponding 2D data points from the 3D space. However, we also need to define the perceived probability before drawing the sensing map in the 2D detection area.

$$P_{ij} = \begin{cases} 1/\alpha e^{\beta d_{ij}} & (\lambda \leq a) \\ 0 & (\lambda > a) \end{cases} \quad (4)$$

Here, $a$ is the variable cost factor and $P_{ij}$ is the probability of point $j$ being detected by the sensor node in point $i$; $\alpha, \beta$ are the probability parameters; and $d_{ij}$ is the Euclidean distance between the points $i$ and $j$.

The perceived probability model for the deployment of sensor nodes should depend on the actual applications of WSNs. The detection area of a node in the Boolean sensing model is a sphere, where the center of circle is the node and the radius is the farthest detection area. If a point lies within this sphere, the probability of detection is 1, otherwise, the probability is 0.

However, the model is often subject to the actual terrain, the distance between nodes and detected points, the number of neighbors for the nodes, the cost factor, and so on. That is why we use the model defined in (4).

*3.4 Sensing-map*

As described in the study by Onur et al. (2010), we can produce the sensing map in 2D space if the detection probability is regarded as the altitude of the 3D terrain.

$$P_j = 1 - \prod_{i=1}^{R}(1 - P_{ij}) \quad (5)$$

The data points in the surveillance area will be detected by more than one sensor node. Here $P_j$ is the detection probability of the point $j$ lying within the detection areas of $R$ sensor nodes.

$$rate = s_1/s_2 \quad (6)$$

The network coverage ratio is one of the standards that measure the performance of WSNs, and it describes a ratio between $s_1$, the number of data points whose detection probability surpasses the threshold; and $s_2$, the total number of data points in the surveillance area. Here $P_t$ is the threshold that reflects the coverage.

In Figure 8, we present the sensing map with a 2D matrix, which has $L \times W$ detection probabilities, according to (5). Twenty sensor nodes are deployed in the Figure 8. The sensing map is non-continuous.

**Figure 8** Sensing map

*3.5 Watershed Transformation*

The watershed algorithm (Vincent and Soille, 1991) is a recently developed image segmentation method based on mathematical morphology. It is also a classical description of topography and is mainly applied for image transformation. For example, the Rocky Mountain watershed divides the United States into two regions. A drop of water flows into the Atlantic if it falls on one side of this watershed; otherwise, it will flow through the other side into the Pacific. The Atlantic and the Pacific are the corresponding lowest regions of these two basins.

The basic idea of watershed transformation is generated from the line reconstruction of geodesy. The watershed is applied into the image processing and the immersion-simulation process of watershed is proposed. The water begins to rise from the minimum points, which are regarded as the local lowest points. The local minimum points affect the formation of the catchment basin. To prevent water from different catchment basins from merging together, dams are built in all confluences as the water lever rises and the watershed is produced after the water lever stops rising.

**Figure 9** Catchment basin

In this study, the missed-detection probability is obtained from the sensing-map by

$$P_{miss} = 1 - P_j \quad (7)$$

The catchment basin is produced by the missed-detection probability, where every gray-scale value represents the corresponding altitude, as shown in Figure 9.

The sensor nodes are placed on the minimum location of the catchment basin. The water starts to rise from the minimum of the gray image. The dams can be built as the water rises from the different basins, and watershed contours shown in Figure 10 are produced.

**Figure 10** Water contours

As shown in Figure 10, the water contours are produced from the sensing-map described in Section 3.4 by the watershed transformation. It represents a series of grid points that lie far away from the sensor nodes within the available detection area of nodes. For example, the grid points in the contours *A*, *B*, and *C* are less likely to be detected by the sensor nodes. As long as the grid points in the water contours can meet the coverage requirements, the others can also do so. Moreover, many paths are produced by the contours. Therefore, in the section below, we discuss how to find an optimal breath path composed of water contours.

## 3.6 Finding optimal breath path

The detection probability, which is called the optimal coverage probability from the optimal breath path, is regarded as one of the standards to judge the coverage of networks. This so-called optimal breath path is the path through which a mobile object can pass, from the perspective of the object, to the other edge of the entire detection area with a poorer detection probability. In this study, the path along which an object can start out from one edge to reach the other edge consists of a series of points that can minimize the sum of all the detection probabilities of data points. From the object's point of view, they do not expect to be detected by the sensor nodes, which is obviously not our aim to gain a better coverage.

The $L \times W$ data points that are present in the 2D plane obtained from the 3D space can maintain the topology of the 3D terrain. The breath path extracted from all the water contours is composed of a series of data points (Onur et al., 2004) derived from the $L \times W$ 2D points. Thus, $PS = [local\_1 ... local\_R]$, where $R$ is the number of points in the optimal path.

**Figure 11** Longitudinal penetration (dotted line) and cross penetration (solid line)

$$P_{opt} = \max_{\{s=1.2..R\}} P_{PS[s]} \quad (8)$$

$P_{opt}$ is the maximum detection probability of the data points on this path because it can truly reflect the deployment of sensors in the surveillance area adjacent to the path to achieve complete deployment. The boundary region is not included in our analysis. We aim to find the paths that penetrate the entire surveillance area, by including longitudinal penetration and cross penetration.

As shown in Figure 11, many paths are extracted from the water contour, including the longitudinal and cross paths. For example, a mobile object penetrates the region from edge 1 to edge 2, whereas another one runs through this region from edge 3 to edge 4. The same penetration is available to objects in the opposite direction.

**Figure 12** Case 1: (a) no longitudinal path; Case 2: (b) no cross path

## 4 Algorithm description and implementation

The sensing map produced by the perceived probability model and the cost parameters in the process of dimensionality reduction from 3D to 2D space is often not smooth. The case that no paths exist to penetrate the surveillance area within the water contours may occur, as shown in Figure 12. We propose an improved algorithm based on the shortest path to find the optimal breath path and the structure *local*, such as *local(i).x*, *local(i).y*, *local(i).pro*, and so on, maintains the locations and weights of data points. The detection probability defines the points' weights. The structure *total* describes all the breath paths, such as *total(j).s*, *total(j).d*, *total(j).ds*, *total(j).paren*, which are the start, end, the sum of the weights, and the precursor matrix of the *j*th path, respectively. Two cases are as follows: 1) As shown in Figure 12(a), no points belonging to the water contours lie on the edge 1, so no path exists; and 2) The beginning is the edge 3 with no end reaching the edge 4, as shown in Figure 12(b). Clearly, we cannot complete the penetration in both cases.

The workflow of longitudinal penetration algorithm is described below:

Begin

Step 1: $i = 0$, $qq = 1$, $flag = 0$, the *state* of all points is 0;

Step 2: choose the edge 1 with $M$ points and define the *end* of the path, $end = 1$;

Step 2.1: $i = i + 1$, traverse *point i* in edge1 and if $flag = M$, jump to step 5; else, check out whether it belongs to the water contour; if not, $flag = flag + 1$, then jump to 2.1; else, continue;

Step 2.2: mark this *start* point as $A$ and *state(A)* is 1, create a weight matrix *distance* and initialize it; if the point is the neighbour of $A$, the *distance* is the weight; else the *distance* is 65535;

Step 2.3: search the smallest sum of weights in *distance* while the point is on the water contour; if *state* of this point is 0, then mark $B$ and *state(B)* = 1; traverse all points adjacent to $B$; if the sum from $A$ to a point $C$ through $B$ is smaller than the distance $(C)$, then update the *distance*, and $B$ is the precursor point of point $C$.

Step 2.4: for other points, if there still exists points whose *state* is 0, jump to step 2.3; else, continue;

Step 3: *totalcost*=65535, mark the point with the smallest sum of weights while the *distance* is not equal to the *totalcost* as the *end* of a path among the points in edge 2;

Step 3.1: if *end* = 1, then $qq = qq - 1$, and jump to step 3.3; else, continue;

Step 3.2: utilize the structure *total* to retain the *start* point and *end* point.

Step 3.3: $qq = qq + 1$; if $i \neq M$, jump to step 2; else, continue;

Step 4: if $qq > 1$, find the smallest sum of weights from $qq - 1$ paths, then $P_{opt1} \leftarrow max\{P_i\}$, and the output is $P_{opt1}$; else, continue;

Step 5: traverse all points in the water contours, if the *extent* of points surpasses 2, give all contours adjacent to this point a weight labelled by the biggest weight in this contour, respectively, else, jump to step 7;

Step 6: find the smallest weight of contours and mark this weight to $P_{opt1}$; the output is $P_{opt1}$;

Step 7: mark the biggest weight of all points in the water contours as $P_{opt1}$, then export it;

End

In this algorithm, the primary distance matrix (Weiss, 1999) represents the sums calculated by all points in the surveillance area and the terminal distance matrix contains the smallest sums of the weights from a point to all other points. The computational complexity is $O(n^2)$, where $n$ is the number of data points, and the cross penetration, whose output is $P_{opt2}$, is similar to the longitudinal penetration and we label the smaller output as the optimal coverage probability $P_{opt}$.

## 5 Simulation results and analysis

In this study, our simulations are based on the *MATLAB R2010a* platform and the coverage parameters are given in Table 1. A 3d Gaussian terrain is thus generated where $\rho = 1000$, $\eta = 5$, $K=20$ and $1 < \omega < 100$. Herein, $\omega$ ranges from 1-100, with the uniform distribution. The projection region in the 2D plane is a rectangular area of $50 \times 50\ m^2$, without consideration of the impact of the border. The sensor nodes are placed with uniform random distribution in the detection area. When we carry out the simulation experiments, different values of the dimensionality reduction parameter $k$ and the detection probability parameters $\alpha, \beta$ are applied to analyze the optimal detection probability and coverage ratio, with an appropriate cost parameter. The threshold of perceived probability is $P_t = 0.80$; $P_{thed}$ is the threshold of coverage ratio and $P_{thed} = 0.90$. Then we comprehensively analyze the impact of the various parameters on the essential sensor nodes.

**Table 1**   Coverage parameters

We have applied different probability parameters, which are subject to the cost factor, to calculate the optimal probability. The cost factor $a = 4$, probability parameter $\alpha = 3$, and the reduction parameter $k = 6$. Each experiment has been repeated 20 times, and the average is the final result. The results are depicted in Figure 13. As the parameter $\beta$ increases, the corresponding detection probability gradually becomes smaller, which means that the perception ability of the deployed nodes is weaker and that more nodes are needed. The coverage ratio decreases with the increase in the parameter $\beta$. The coverage ability and reliability of the networks become worse in a bad environment.

**Figure 13**   Effect of the perceived parameter $\beta$ on the detection probability and coverage ratio

Similarly, for $\beta = 0.2$, other parameters are the same as in Figure 13, but the value of $\alpha$ changes. Then the trend of $P_{opt}$ with reference to the coverage parameters is observed, and the detection probability and coverage ratio decrease with a slow change, as in Figure 14. Because $P_{opt}$ in Figure 14 does not change enormously as in Figure 13, compared with $\beta$, the parameter $\alpha$ has lesser influence on the coverage of sensor nodes.

**Figure 14**   Effect of the perceived parameter $\alpha$ on the detection probability and coverage ratio

During the process of nonlinear dimensionality reduction, the reduction parameter $k$ affects the optimal coverage probability and changes the breath path. If we apply a bigger value of $k$, a better convergence will be lost while simultaneously increasing the computational complexity. A smaller value of $k$ leads to fewer neighbors, and the local reconstructed weight matrix may not be able to maintain a more manifold invariant and the topology of data points perfectly. In our simulations, twenty sensor nodes are placed in the surveillance area, with $\beta = 0.3, \alpha = 2$, and $a = 4$ as in Figure 15. The probability $P_{opt}$ increases at the beginning and then decreases as the parameter $k$ changes. Finally, it maintains a fluctuating state.

**Figure 15**   Effect of the reduction parameter $k$ on the detection probability and coverage ratio

The cost parameters affect the perceived probability model that considers the topology of data points in the 3D terrain as a rough terrain. In Figure 14, the detection area is influenced by the 3D terrain, which leads to the blind area where parts of the regions cannot be detected by the sensor nodes. Approximate cost parameters are needed to maintain the topology of the surface and complete the transformation from the 3D space to the 2D plane.

In Figure 16, twenty sensor nodes are placed with a uniform random distribution in the $50 \times 50\ m^2$ area, where $\beta = 0.3, \alpha = 2$, $k = 9$. According to the deployment strategy, the detection probability increases firstly, and then stops increasing with fluctuations rising in certain values.

**Figure 16**   Effect of the cost parameter $a$ on the detection probability and coverage ratio

The sensor nodes can detect objects in a larger area with increases in the parameter $a$, until the detection ability of nodes begins to decline; the cost parameter $a$ has little influence on $\lambda$ and the detection probability does not increase any more. However, we need to consider the locations of the sensor nodes deployed on the surface because points, such as *A* or *B*, whose neighbors may have a larger value for $\lambda$ than those of *C* or *D*, exist. The cost factor $a$ affects the effective detection area of points and makes it relatively smaller. Compared with the Figure 16, the sensor nodes are placed on a steeper surface in Figure 17, which leads to a smaller detection probability and a lower coverage ratio. More sensor nodes are needed to be deployed in Figure 17.

**Figure 17**   Effect of the cost parameter $a$ on the detection probability and coverage ratio for a steeper surface

**Figure 18**   Points *A* and *B* have a smaller cost value than *C*

Because the resources of WSNs are limited and the deployment is influenced by the real-world applications, the coverage strategy we have proposed aims to implement optimal coverage, which needs to cover the surveillance area using as few number of sensor nodes as possible and reduce the energy consumption largely to meet the requirements of WSNs. In Figure 19, the perceived probability parameters, $\beta = 0.3, \alpha = 2$, $k = 9$, and $a = 3$. The detection probability $P_{opt}$ increases progressively with increasing sensor count. The value of $P_{opt}$ is 0.91, which is more than the probability threshold $P_{thed}$. However, the coverage ratio is 0.62 and is less than $P_t$ when the sensor node count is 30. Both the coverage ratio of networks and detection probability can reach the optimal levels if the sensor count is maintained at 40-50.

**Figure 19** Effect of the sensor count on the coverage parameters

The parameters in Figure 20 are the same as in Figure 19, where $\beta = 0.3, \alpha = 2$, $k = 9$, and $a = 3$. The sensor nodes are placed on the surface with uniform random distribution. However, some nodes are placed on the steep surface through the analysis of cost value $\lambda$ so that more nodes are placed under this situation, compared to that in Figure 19, where 40-50 sensor nodes are required to achieve the optimal coverage. However, the coverage ratio is only 0.5. The detection probability $P_{opt}$ also increases progressively with the increasing sensor count and 60-70 sensor nodes are needed to meet the requirements of the optimal thresholds of both detection probability and coverage ratio. In conclusion, an approximate coverage parameter has a crucial effect on the coverage strategy of sensor nodes and the stability of network performance.

**Figure 20** Effect of the sensor count on the coverage parameters for a steeper surface

## 6 Conclusion

In this study, we have proposed a new coverage strategy for nodes available to a 3D complex terrain in the real-world applications. We use coverage parameters, such as the cost factor, reduction parameter, and perceived probability parameters, to complete the transformation from 3D space to the 2D plane and determine the deployment strategy of sensor nodes. We propose an algorithm based on the perceived probability to find the optimal breath path and achieve the optimal detection probability $P_{opt}$ and determine the number of sensor nodes needed to cover the entire surveillance area. Meanwhile, the possible cases occurring in the penetration process are included in our analysis. The experimental simulation results are consistent with the existing theories and all the coverage parameters meet the requirements of WSNs after the deployment, which prove the feasibility of the transformation strategy from 3D to 2D space and the necessity to analyze the real-world applications of WSNs before deployment of sensor nodes. In the future, we will try to make improvements on the 3D terrain with the inclusion of obstacles which can reflect the real applications of WSNs. Moreover, mobile nodes would be included, which will make the coverage issue and deployment strategy more complex. Energy consumption of nodes that have great effect on the performance of WSNs will be also considered.

## Acknowledgement

This work is supported by National Natural Science Foundation of China (Grant No. 61173163, 61202443), Program for New Century Excellent Talents in University (Grant No. NCET-09-0251), and the Fundamental Research Funds for the Central Universities (DUT13JS07).

**Figure 1** Sensor nodes distributed on a regular surface

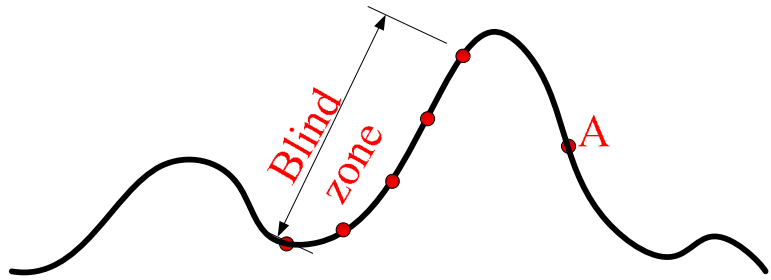

**Figure 2** Actual applications: (a) is the real world terrain, and (b) is the simplified regular surface

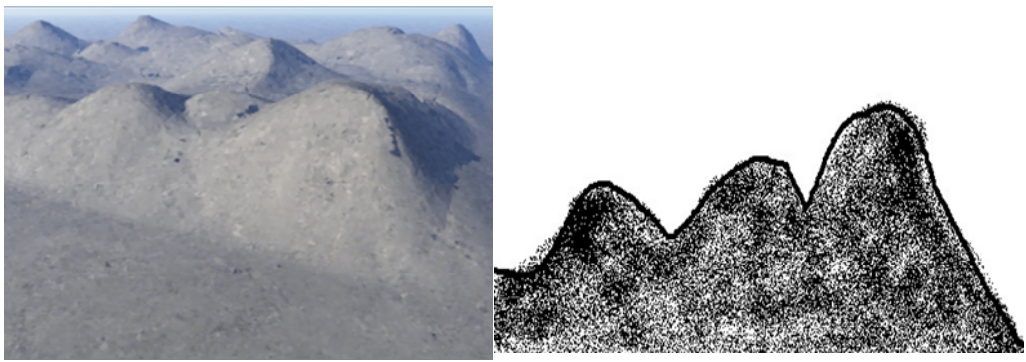

(a)          (b)

**Figure 3** Peak map ($K = 1$, $\omega = 20$)

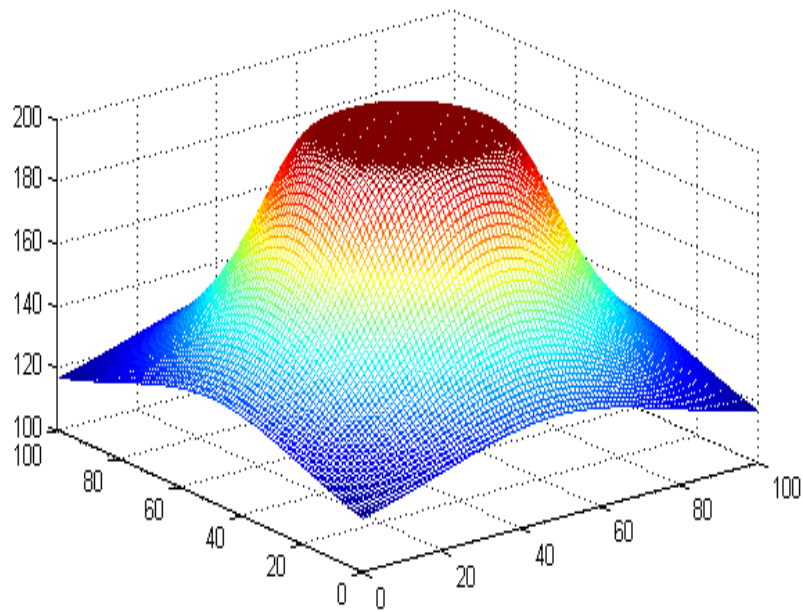

**Figure 4** Three-dimensional terrain ($K=20$)

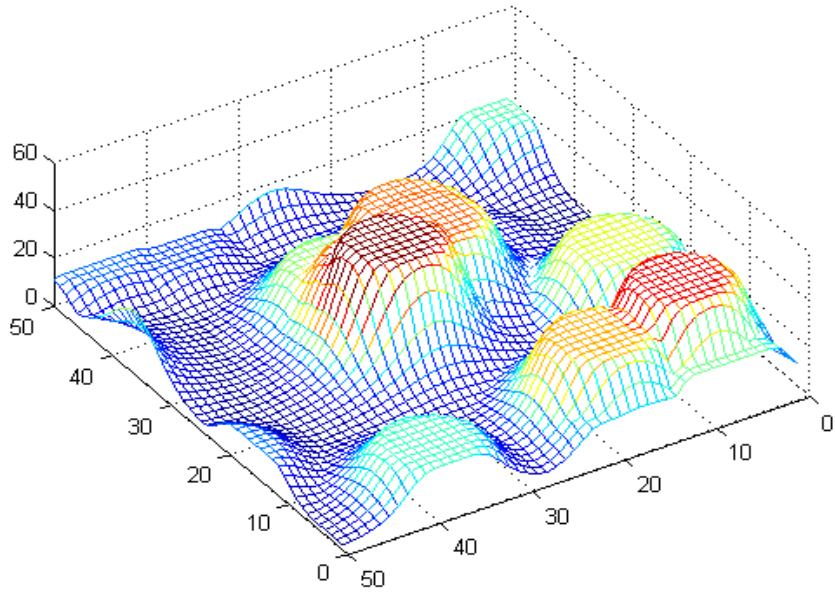

**Figure 5** Comparison between PCA and LLE

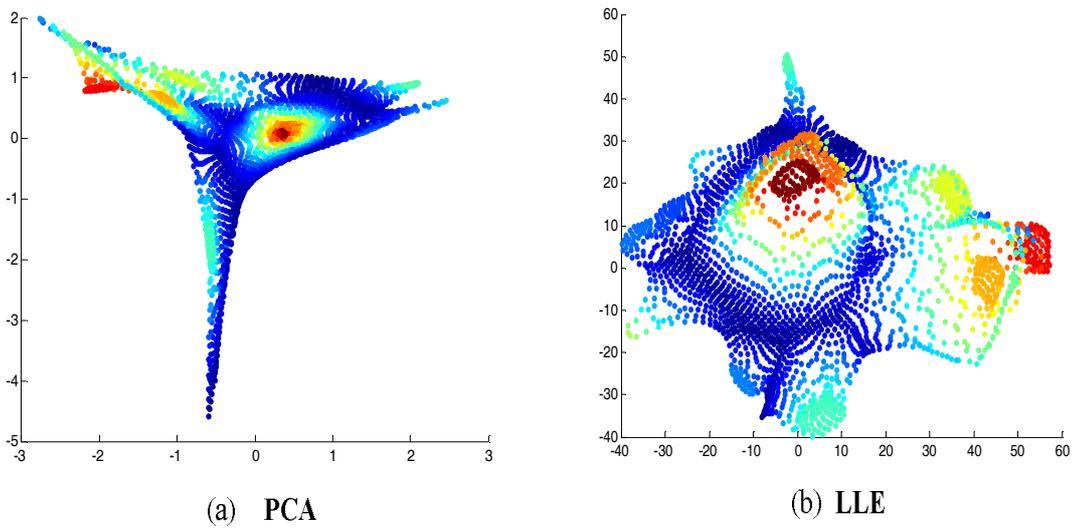

(a)  PCA  (b)  LLE

**Figure 6** Altitude difference of data points on 3D terrains

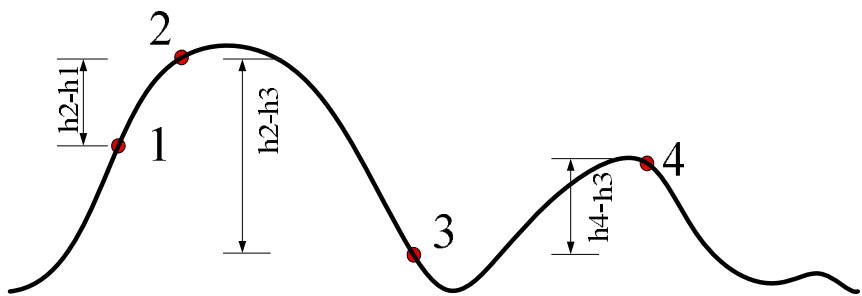

**Figure 7** Euclidean distance between the 2D points

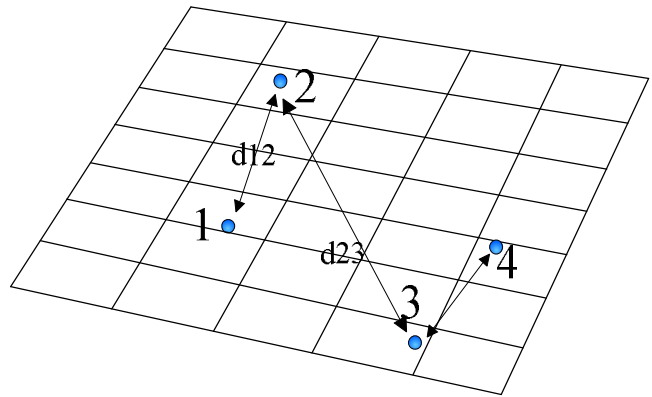

**Figure 8** Sensing map

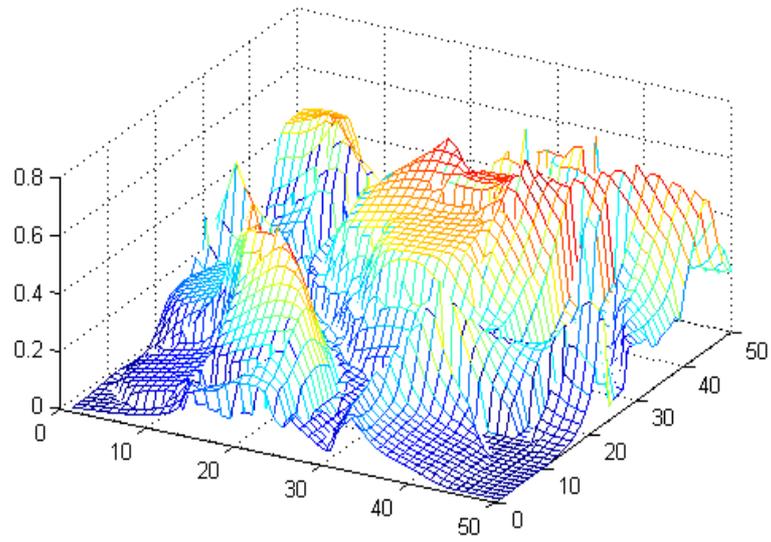

**Figure 9** Catchment basin

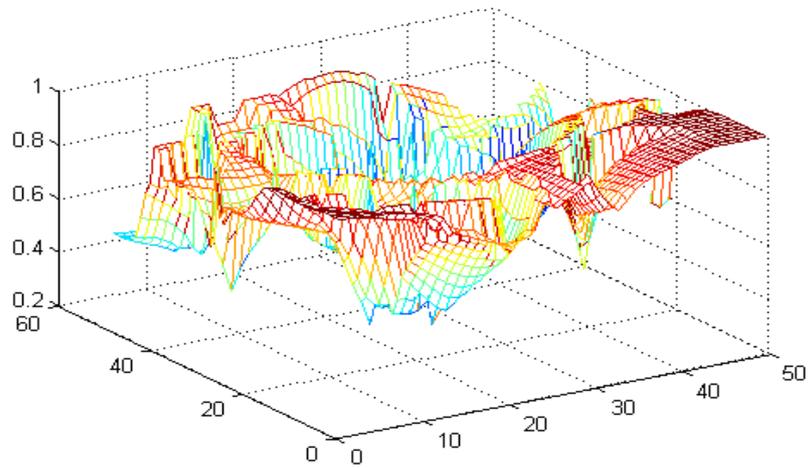

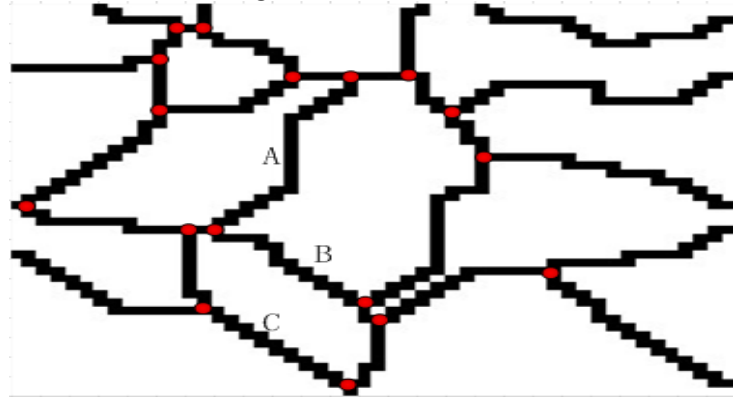

**Figure 10** Water contours

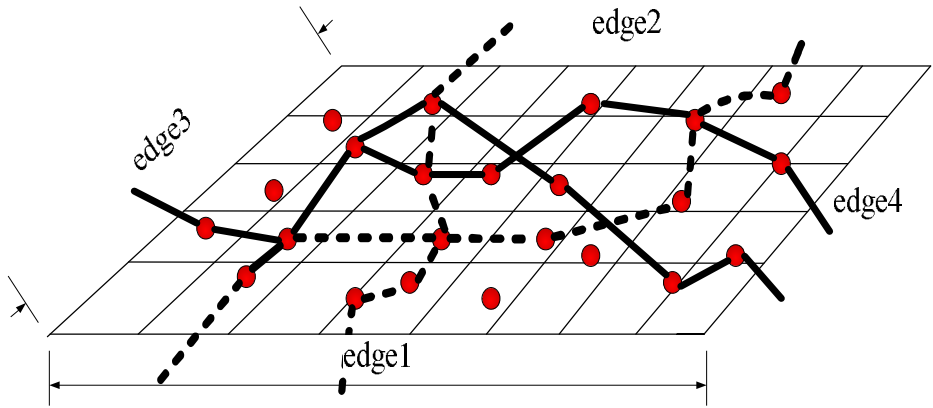

**Figure 11** Longitudinal penetration (dotted line) and cross penetration (solid line)

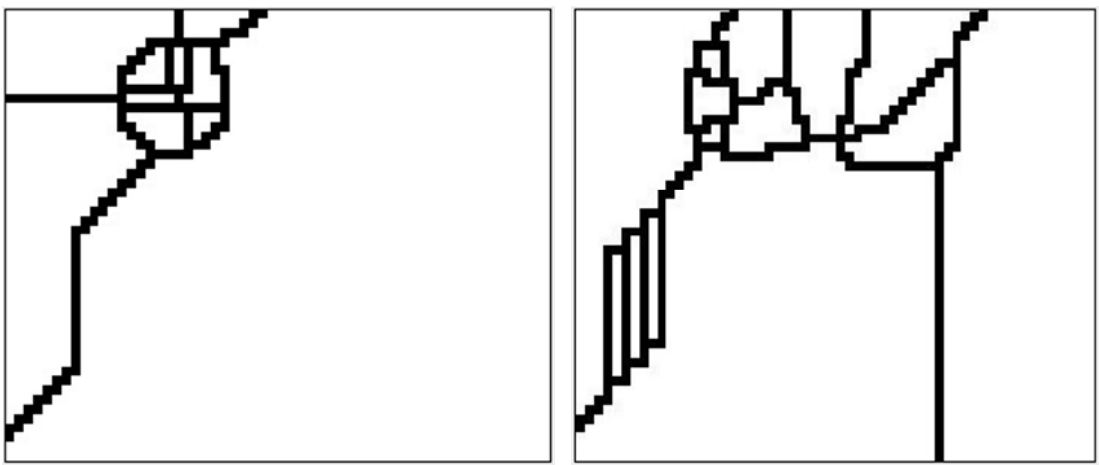

**Figure 12** Case 1: (a) no longitudinal path; Case 2: (b) no cross path

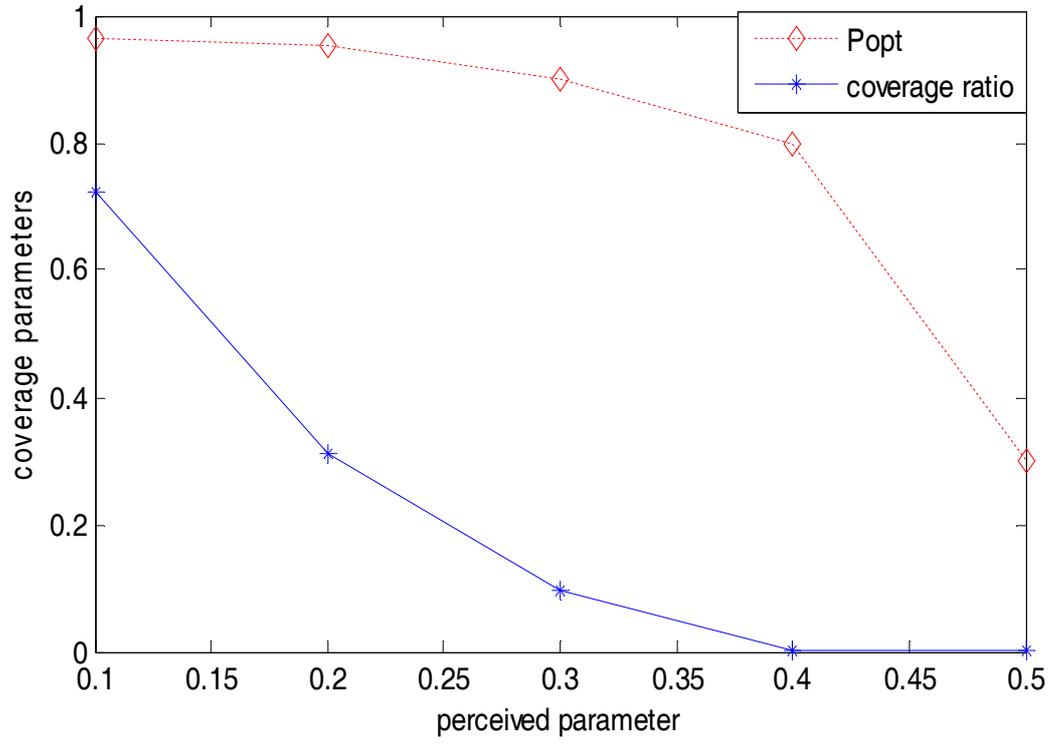

**Figure 13** Effect of the perceived parameter $\beta$ on the detection probability and coverage ratio

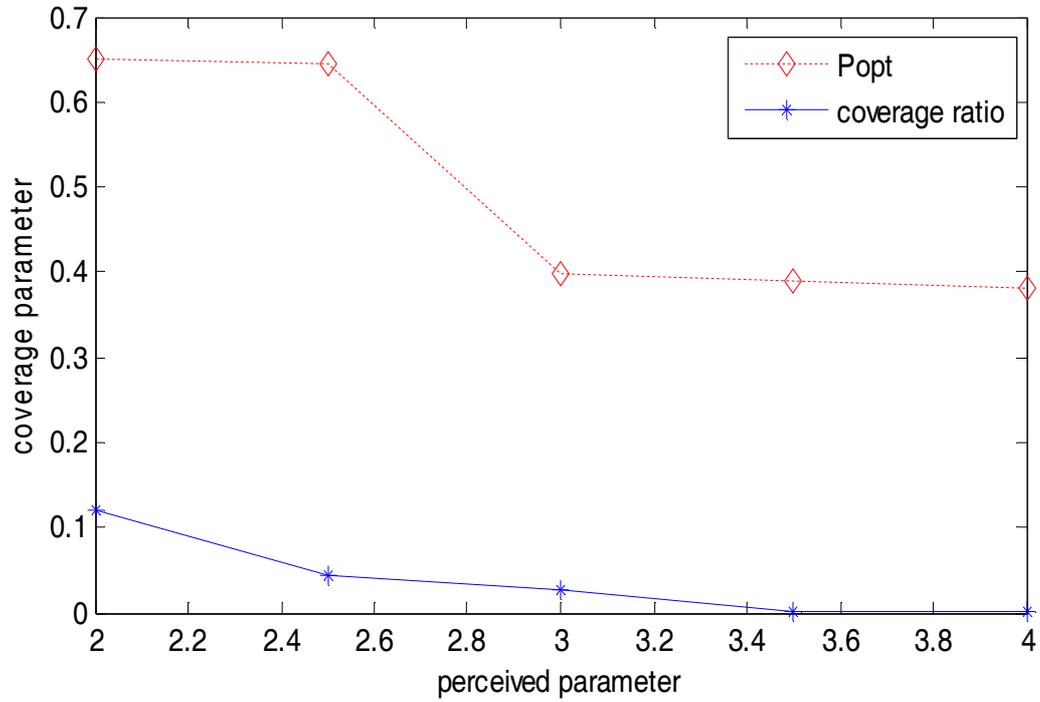

**Figure 14** Effect of the perceived parameter $\alpha$ on the detection probability and coverage ratio

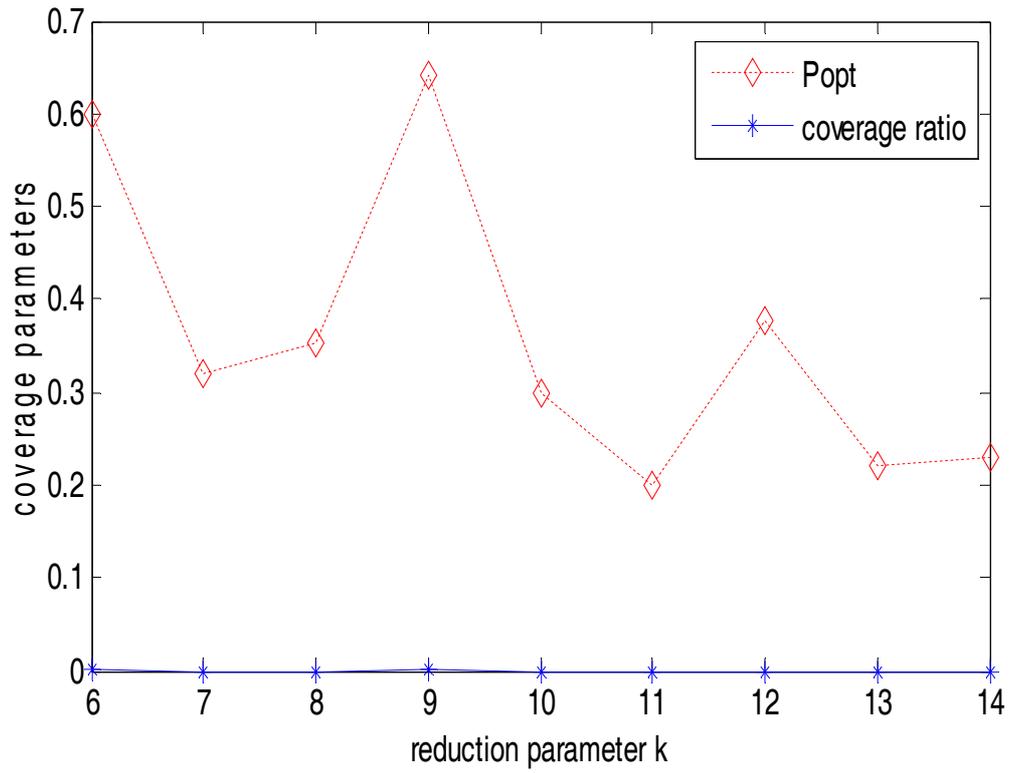

**Figure 15** Effect of the reduction parameter $k$ on the detection probability and coverage ratio

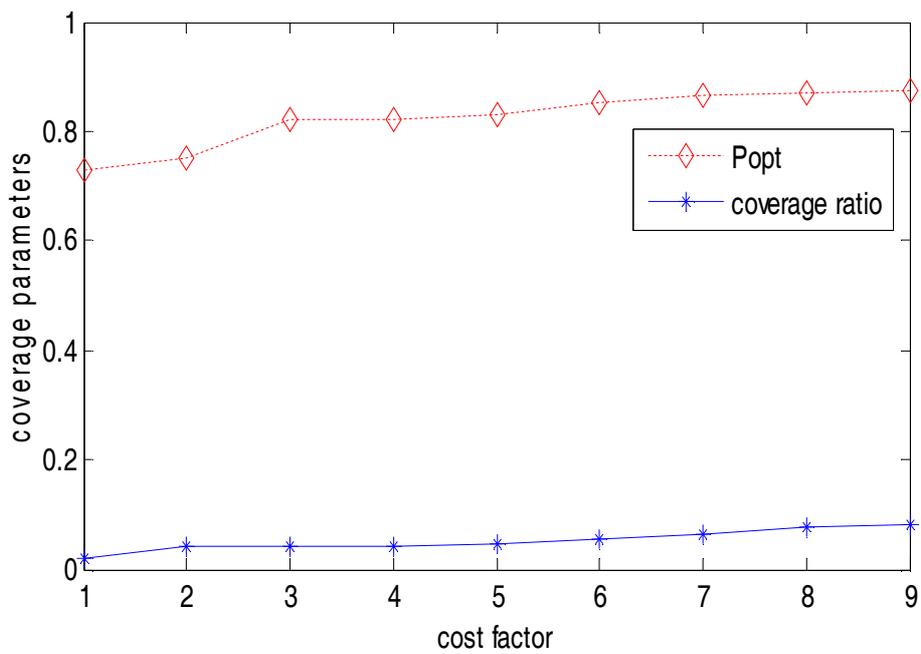

**Figure 16** Effect of the cost parameter $a$ on the detection probability and coverage ratio

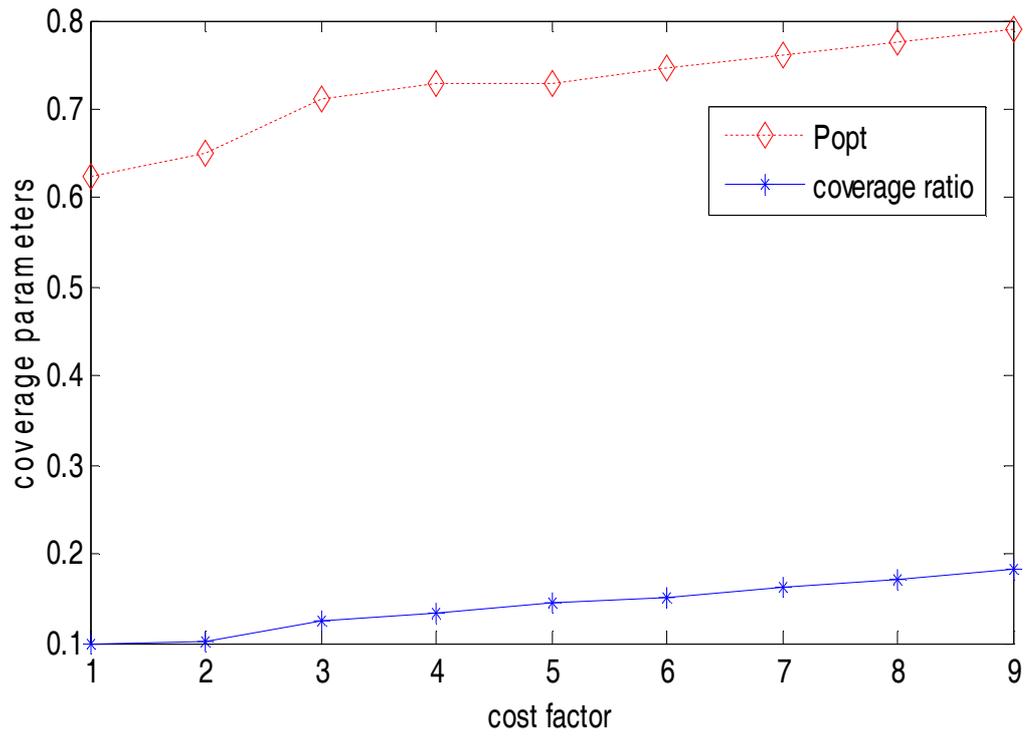

**Figure 17** Effect of the cost parameter $a$ on the detection probability and coverage ratio for a steeper surface

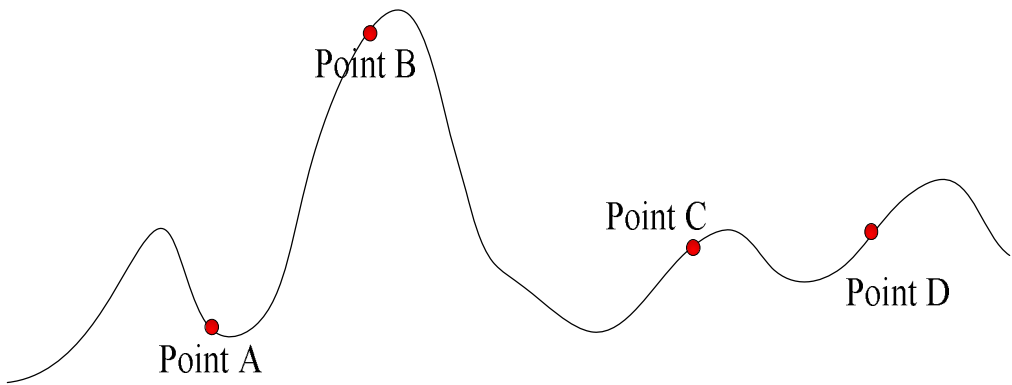

**Figure 18** Points $A$ and $B$ have a smaller cost value than $C$

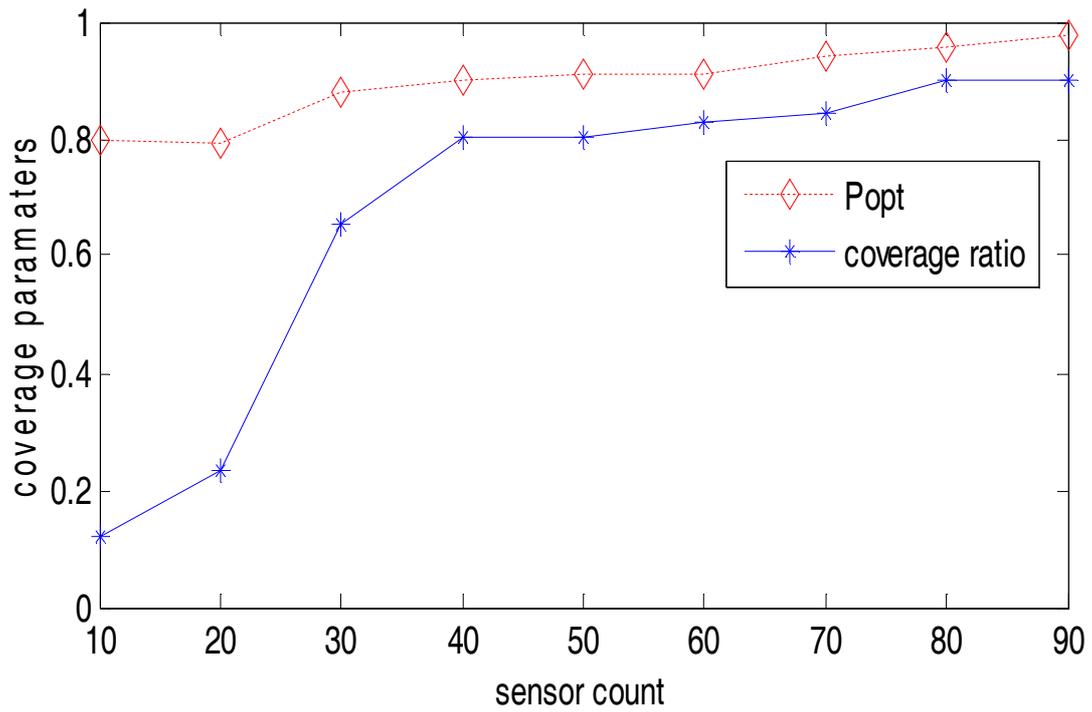

**Figure 19**  Effect of the sensor count on the coverage parameters

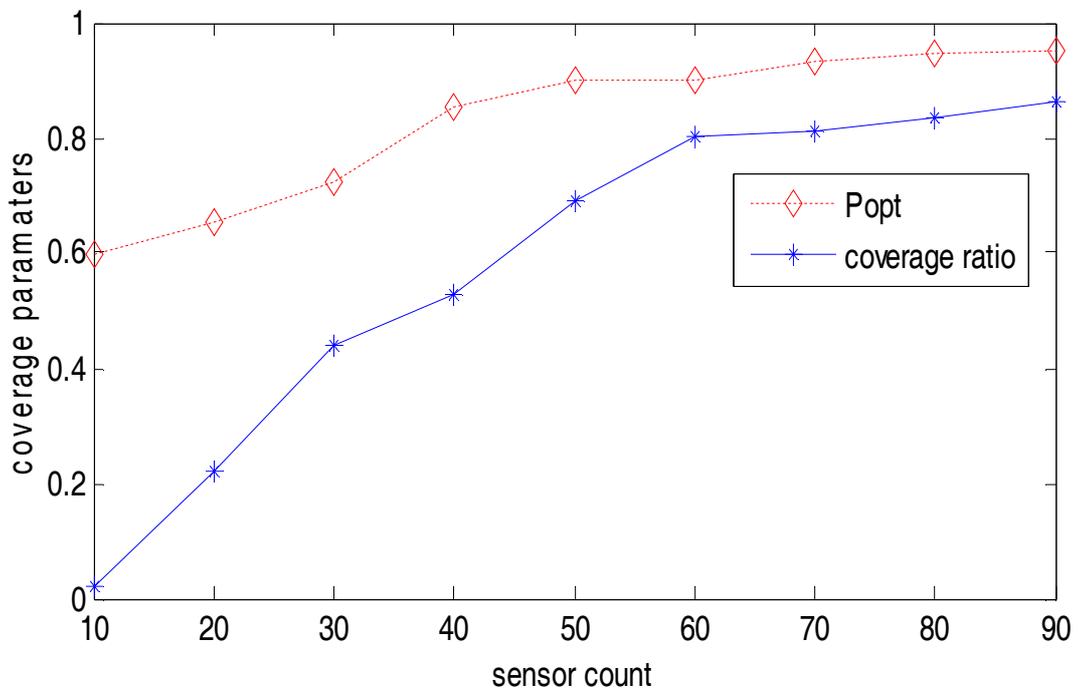

**Figure 20**  Effect of the sensor count on the coverage parameters for a steeper surface

**Table 1** Coverage parameters

| PARAMETER | VALUE | PARAMETER | VALUE |
|---|---|---|---|
| $L$ | 50 | $\eta$ | 4 |
| W | 50 | $\alpha$ | 2~4 |
| K | 1~100 | $\beta$ | 0.1~0.5 |
| $\rho$ | 1000 | $d_r$ | 10 |
| $P_t$ | 0.80 | $P_{thed}$ | 0.90 |
| $\omega$ | 1~100 | a | 1~9 |
| $k$ | 6~14 | | |